# Deep Learning-Based Automatic Delineation of Liver Domes in kV Triggered Images for Online Breath-hold Reproducibility Verification of Liver Stereotactic Body Radiation Therapy

Sugandima Weragoda[1,2], Ping Xia[2], Kevin Stephans[2], Neil Woody[2], Michael Martens[1], Robert Brown[1], Bingqi Guo[2]

Cleveland Clinic Foundation, Cleveland, OH[1].

Case Western Reserve University, Cleveland, OH[2].

Abstract

Background: Stereotactic Body Radiation Therapy (SBRT) can be a precise, minimally invasive treatment method for liver cancer and liver metastases. However, the effectiveness of SBRT relies on the accurate delivery of the dose to the tumor while sparing healthy tissue. Breath-hold techniques, together with spirometry devices, are used to reduce liver motion. However, challenges persist in ensuring breath-hold reproducibility, with current methods often requiring manual verification of liver dome positions from kV-triggered images. To address this, we propose a proof-of-principle study of a deep learning-based pipeline to automatically delineate the liver dome from kV-planar images, thereby enhancing accuracy and reducing manual intervention in SBRT treatments.

Purpose: Fast, accurate, and automatic liver dome delineation on kilo-voltage trigged images can assist online breath-hold reproducibility verification for liver stereotactic body radiation therapy (SBRT) as proposed in a previous study[1] from our institution. Using a deep learning-based pipeline, we conducted a proof-of-principle study for this purpose.

Methods: From 24 patients who received SBRT for liver cancer or metastasis inside liver, 711 KV-triggered images acquired for online breath-hold verification were included in the current study. We developed a pipeline comprising a trained U-Net for automatic liver dome region segmentation from the triggered images followed by extraction of the liver dome via thresholding, edge detection, and morphological operations. The performance and generalizability of the pipeline was evaluated using 2-fold cross validation, in which the patient dataset was split into two folds, with $Fold_1$ used as the training dataset and $Fold_2$ as the test dataset initially and then vice versa. We used two metrics for evaluation of the pipeline: Root Mean Square Error (RMSE) between the predicted liver dome and ground truth based on manual delineation, and rate of dome detection per patient.

Results:  The training of the U-Net model for liver region segmentation took 30 minutes on an augmented training data set containing 500 images. The end-to-end automatic delineation of a liver dome for any triggered image took less than one second. The RMSE and rate of detection for $Fold_1$ with 366 images was (6.4±1.6) mm and 91.7%, respectively. For $Fold_2$ with 345 images, the RMSE and rate of detection was (7.7±2.3) mm and 76.3% respectively.

Conclusion: A deep learning-based automatic segmentation pipeline was proposed for liver dome delineation in kV triggered images acquired during SBRT treatments. This study proved the feasibility of fast and accurate segmentation of liver dome from kV X-ray images.

## Introduction

Surgical resection is considered the gold standard for treating liver cancer and liver metastases. For patients who are ineligible for surgery, Stereotactic body radiation therapy (SBRT) is gaining traction as a minimally invasive, ablative technique for primary and metastatic liver tumors[2]. SBRT is based on the principle of delivering a highly conformal dose distribution with a rapid dose drop-off, making it possible to avoid irradiating healthy regions of the liver while escalating the dose to the treatment area. However, for SBRT to be effective, the dose delivery to the tumors must be precise so that the remaining healthy liver tissue are protected. This requires strategies to mitigate motion effects such as breathing that could result in uncertainty in dose delivery and target dose coverage.

Spirometry devices such as Active Breathing Coordinator, have been used for breath-hold liver SBRT to minimize breathing motion[3]. However, the reproducibility of every breath-hold for patients under breath-hold treatments is challenging. Many studies have been conducted on breath-hold reproducibility using CT[4,5], planar images[4,6,7], fluoroscopy[6], and CBCT[8–11], and non-radiation imaging techniques[12–15]. Eccles et al.[4] study breath-hold reproducibility with ABC and highlight the need for additional image guidance to maintain accuracy. Lu et al.[5] demonstrate that poor breath-hold reproducibility may necessitate acquiring multiple CTs during simulation to ensure the robustness of treatment plans. Dawson et al.[6,7] reveal how daily image guidance substantially improves setup accuracy compared to using ABC alone. Zhong et al.[8] note that while long breath-holding techniques reduce liver motion, significant errors persist, which can be mitigated via CBCT-guided online corrections. Hawkins et al.[9] also show that CBCT can improve the accuracy of liver positioning, reduce setup errors, and enhance treatment precision. Kawahara et al.[10,11] note that diaphragm matching techniques for positioning are less error-prone compared to bone matching techniques and suggest this method for breath-hold liver SBRT. Naumann et al.[12] report that integrating optical surface guidance with ABC enhances liver target positioning, and Bloemen-van Gurp et al.[13] that integrating ultrasound reduces uncertainty in target localization. Additionally, Weykamp et al.[14] and Feldman et al.[15] show promising results with MR-guided SBRT, suggesting reduced liver doses and milder toxicities. These studies conclude that for filtering out non-compliant breath-holds, matching the liver dome from the planning CT with every trigger image acquired during treatment prior to each breath-hold is recommended.

The results of a previous study from our institution which used kV triggered images taken at the beginning of each breath hold to verify the breath-hold reproducibility concluded that visually verifying the liver dome position from triggered images eliminated breath-holds with >10 mm deviations from the planned dome position and reduced breath-holds within 5mm-10 mm deviations from 15% to 11%[1]. Automatic delineation of the liver dome and subsequent automatic beam gating would improve the accuracy of this procedure with less subjectivity or human intervention.

Most conventional segmentation methods for liver domes are from volumetric images such as CT and are based on models incorporating prior knowledge of the liver dome anatomy[16–19] which are not applicable to the planar triggered images. The major challenges in automatic delineation of liver domes in triggered images include: (A) the visibility of liver dome in triggered images may be low due to poor soft tissue contrast; and (B) liver dome position and shape in the triggered images depend on the patient anatomy, location of the isocenter, and projection angles of the kV triggered images

To address these challenges, we propose a deep learning-based (DL-based) pipeline. We employed a U-Net, which is a DL architecture specifically designed for biomedical image segmentation. We used initial weights of a U-Net pretrained on ImageNet[20], thereby leveraging transfer learning capabilities for improved performance[21]. We also utilized the HED[31] model which is a DL model pre-trained to detect natural edges of objects in images. The incorporation of the HED[22] model in DL-based medical image segmentation has shown promise[32-35].

Inspired by relevant literature, we propose a fast automatic liver dome delineation pipeline that comprises a trained U-Net, a pre-trained HED model, and morphological operations. To our knowledge, this study introduces the first DL-based liver dome delineation from kV planar images, contributing as a baseline in the literature. Further, we demonstrate a novel use of the pre-trained HED model in medical image segmentation. Lastly, our proposed pipeline has potential use for automatic verification of the breath hold reproducibility for SBRT in liver and abdominal tumor treatments.

## Materials and Methods

### Data

The dataset for this study is taken from a previously published IRB-approved study[1]. The dataset contains 711 kV-triggered images from 24 liver patients treated with SBRT from May 2019 to February 2022 at our institution. Liver dome was manually delineated from the triggered images as the ground truth.

### Deep learning pipeline

The deep-learning-based pipeline was developed in Python using the TensorFlow[23] 2.x library. All implementations were performed on an NVIDIA Quadro-RTX-6000 GPU.

Figure. 1 provides a broad overview of the pipeline. All the triggered images and the manually delineated liver domes from the triggered images were preprocessed and used as inputs for U-Net. The U-Net outputs a predicted liver dome region map which is then postprocessed to extract the liver domes.

The triggered images were divided into two folds with 366 images from 12 patients in $Fold_1$ and 345 images from 12 patients in $Fold_2$. Using these two folds, we conducted two-fold cross validation of the neural network and the pipeline to evaluate the generalizability of the proposed method. The following subsections describe the three stages in Figure 1 in detail: image preprocessing, training of the U-Net, and postprocessing of the liver region predictions.

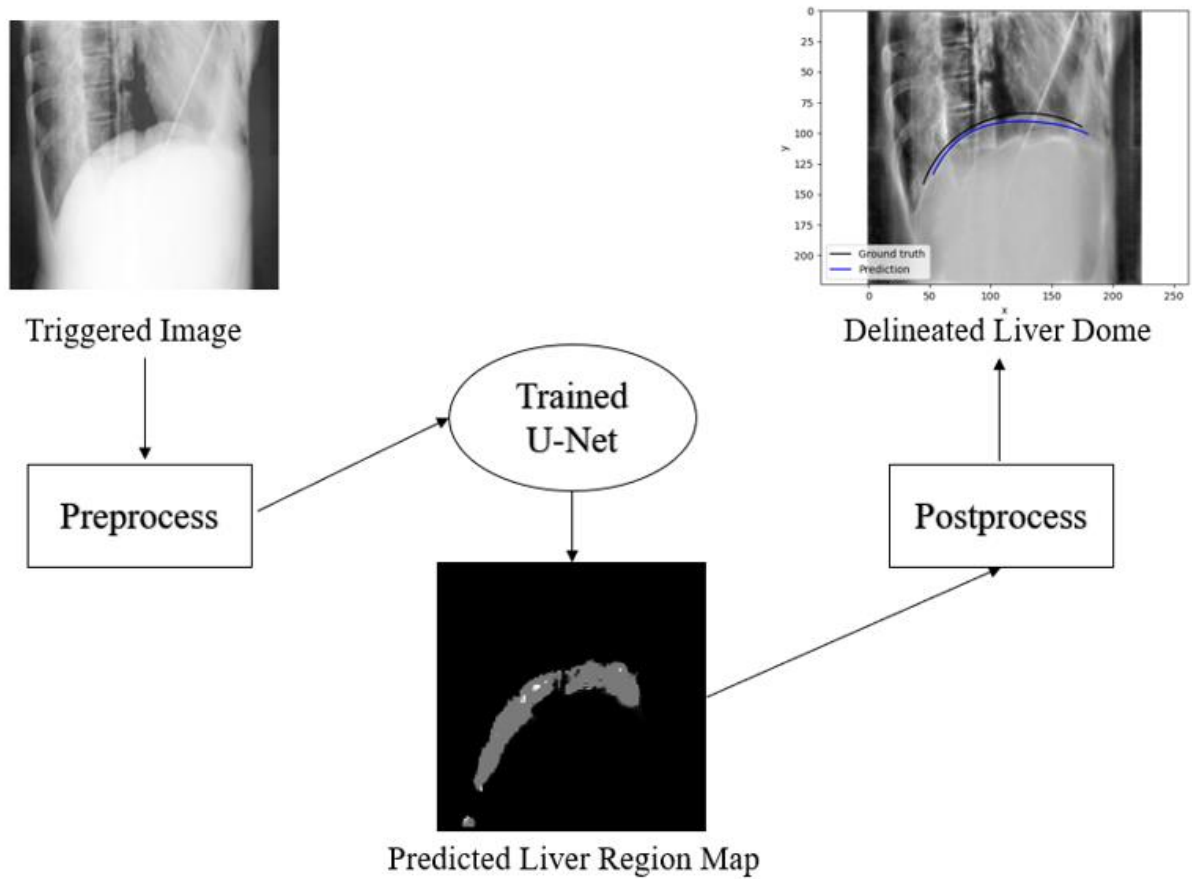

*Figure 1. The proposed deep learning-based pipeline for the delineation of liver domes from triggered images: the triggered image is first preprocessed and then passed into the trained U-Net, which outputs the predicted liver region map. The predicted map is then postprocessed to extract the final liver dome delineation.*

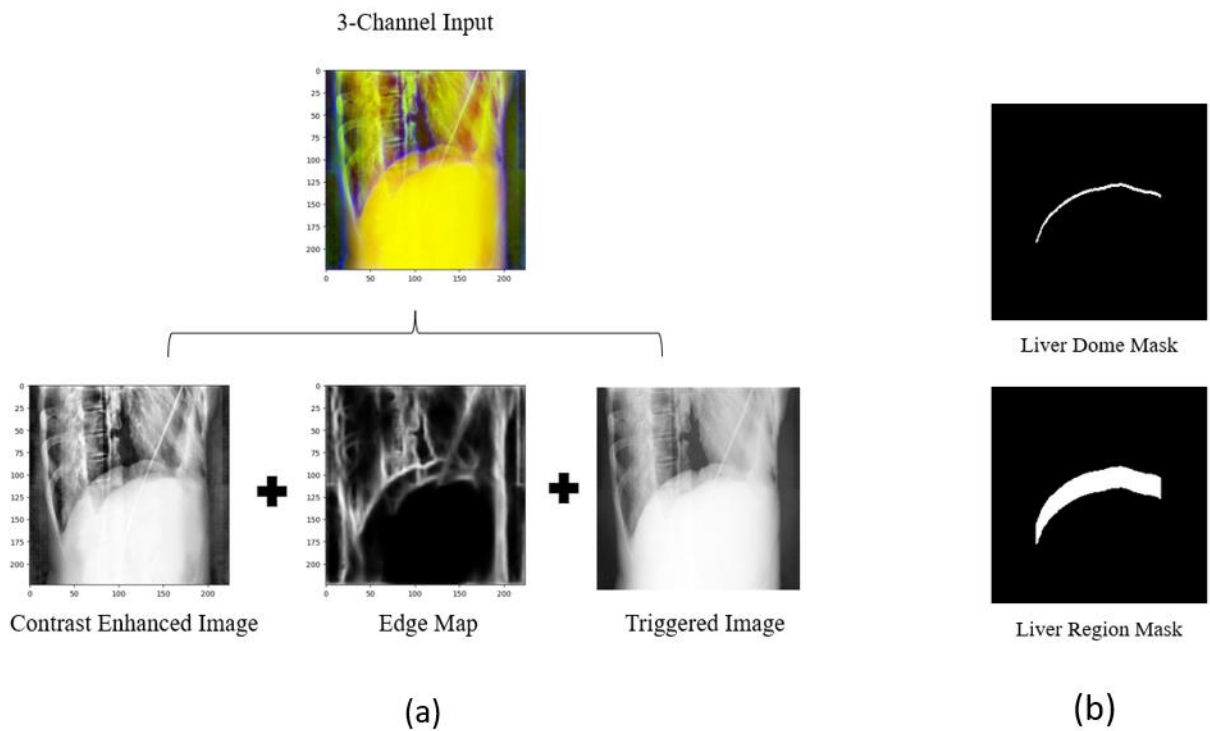

*Figure 2. (a) Preprocessing steps to create the 3-channel input: creating the contrast enhanced image, the HED edge map, and combining them together with the original triggered image to generate the 3-channel input. (b) Manually contoured liver dome ground truth mask and liver region ground truth mask of a triggered image.*

a. Image Preprocessing and ground truth labeling

All the triggered images were subjected to the preprocessing steps depicted in Figure 2(a). The first step of preprocessing was to resize the triggered images from 512x512 to 256x256. Downsizing the images reduce the training time required for the U-Net. The preprocessing was performed to generate a 3-channel input image for the U-Net deep learning model. The 3-channel input consisted of the contrast enhanced image, the edge map, and the original grey scale triggered image.

A kV-triggered image typically suffers from poor contrast, which makes soft tissue differentiation difficult. We chose the python OpenCV implementation of CLAHE[24] (Contrast

Limited Adaptive Histogram Equalization) to enhance the image contrast. The CLAHE is an Image Equalization method developed for low contrast medical images. It is a variation of Adaptive Histogram Equalization[25] (AHE), but mitigates the noise amplification problem prevalent in AHE, by performing histogram equalization in small patches with high accuracy and contrast limiting. The contrast enhanced triggered image was used as the first channel of the 3-channel input.

For the second channel of the 3-channel input, we used the edge map of the triggered image. The edge map was generated by using the Caffe implementation of the pre-trained HED model[26]. The pre-trained HED model was able to predict the edge map of any given triggered image without the need for further training or tuning.

We used the preprocessing stage to incorporate feature engineering. We recognized that contrast enhancement for soft tissue differentiation and edge information for organ contouring are the most useful input features for the neural network during training with the goal of segmenting the liver dome region within the triggered images. Therefore, to facilitate the model's learning of the features most relevant to liver region segmentation, we combined the contrast enhanced image, the edge map and the original grey scale triggered image into a 3-channel image and used it as the input to the neural network.

An example ground truth label for training the neural network for this binary semantic segmentation task is shown in Figure 2(b). The liver dome was manually delineated from the triggered images for analysis in a previous study[1]. In this study, the manual contours were converted to binary maps of the liver dome, and then expanded inferiorly 20 pixels to form the masks of the liver dome region, which served as the ground truth labels for training and validating the neural network.

b. U-Net for Liver Dome Region Segmentation

The second stage of the pipeline is the training and validation of a neural network for the segmentation of the liver dome region in any given triggered image.

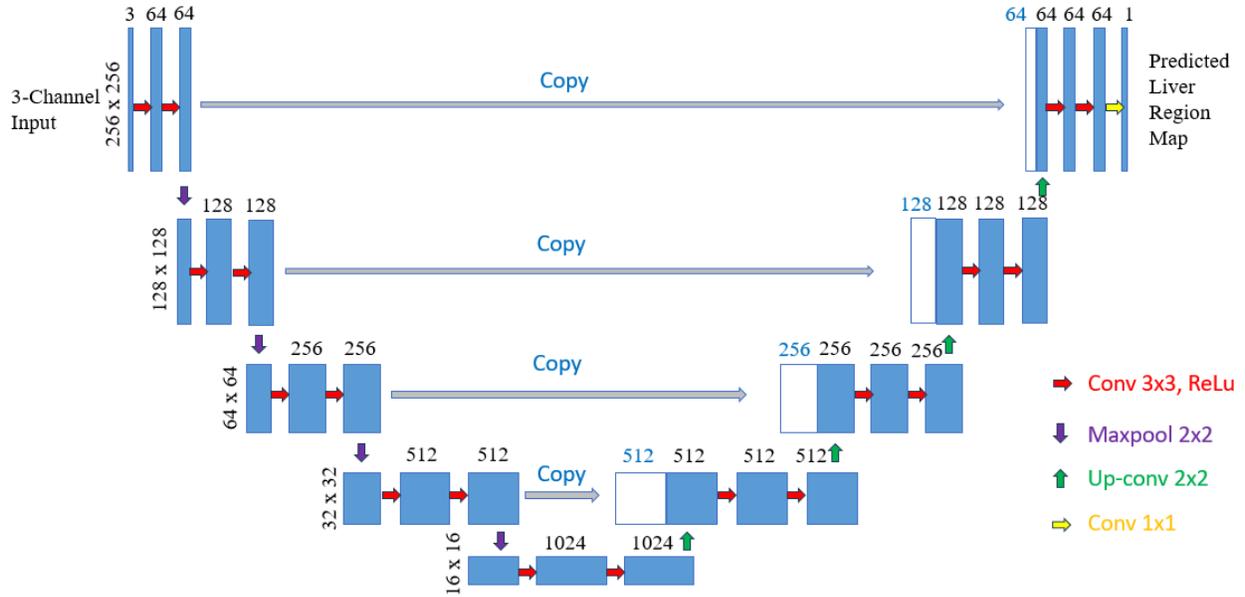

*Figure 3. The architecture of the U-Net model.*

Figure 3 illustrates the U-Net architecture of our neural network implemented in TensorFlow[23]. It consisted of a contraction or encoder path, a bridge layer, and an expansion or decoder path. The encoder path of the U-Net used the technique of transfer learning and consisted of a VGG16 pre-trained on the ImageNet dataset. This portion of the U-Net remained static during the training of the neural net. The remaining layers of the U-Net were trained during model training. The input layer of the model is 256x256x3 to accommodate the 3-channel input images. The output layer is 256x256x1 to produce the desired grayscale map. The model used ReLu[27] activation for all layers except the output layer which used sigmoid activation. To avoid possible overfitting as the training dataset was limited in size, we chose the dropout rates of 0.4, 0.2, 0.2, and 0.1 for the decoder blocks 1,2,3, and 4, respectively.

Two-fold cross-validation was performed to evaluate the generalizability and accuracy of the U-Net. The preprocessed dataset was divided into two folds, each consisting of 12 patients. Each fold contained roughly the same number of triggered images. In the first run, the $Fold_1$ was used to train the U-Net, and the $Fold_2$ was used as a test set to evaluate the accuracy of the segmentations for the liver dome regions. In the second run, the $Fold_2$ was used to train the U-Net, while the $Fold_1$ was used as the test set.

We used data augmentation on each fold to increase the limited training data to 500 images. The augmentation included the geometric transformations: translation, and horizontal flipping, and random adjustments to image brightness and contrast. We used the Python library albumentations[28] to implement the on-the-fly augmentations, just before model training.

The model segmentation performance on each fold was evaluated by reporting the standard metric: Intersection over Union (IoU).

$$IoU = \frac{|PM \cap OM|}{|PM \cup OM|} \quad (1)$$

Where PM is the set of pixels in the predicted region and OM is the set of pixels in original ground truth region mask. |PM| and |OM| are the pixel counts of PM and OM. The IoU metric ranges from 0 to 1.0 (perfect match).

The U-Net was trained on each fold for 100 epochs with early stopping chosen by monitoring the validation loss while training. The model validation while training was performed on a randomly selected 10% of the training dataset. The training batch size was set to 20, the base learning rate was set to 0.001, and AdaMax[29] was selected as the optimizer function. The training was terminated when the validation loss reached 0.1.

Training of the U-Net model on each fold took approximately 30 minutes on an NVIDIA Quadro-RTX-6000 GPU.

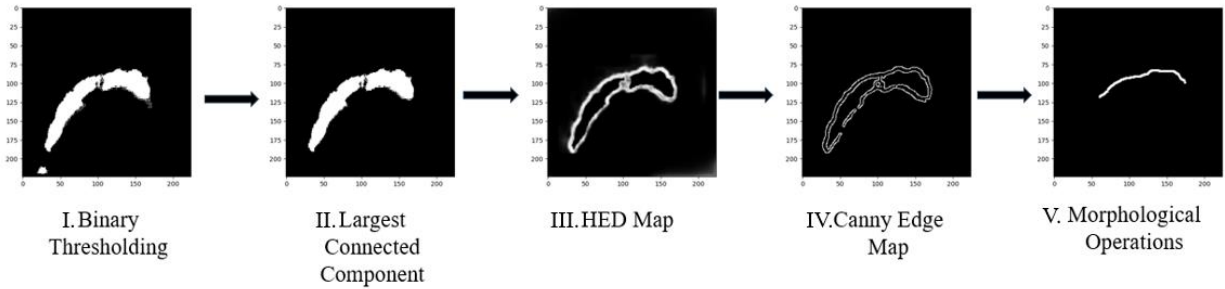

*Figure 4. Postprocessing seps to extract the liver dome from the predicted liver region map. From left to right: I. Prediction map subjected to binary thresholding with threshold set to the image intensity mean, II. Extraction of the largest connected component, III. Application of HED, IV. Application of Canny edge detection, and V. Application of morphological operations.*

c. Postprocessing

The liver dome region segmentations from the U-Net were subjected to postprocessing to extract the liver dome delineations. Figure 4 illustrates the postprocessing steps. All the postprocessing steps were implemented using the OpenCV library in Python.

First, the grayscale prediction maps of the dome region from the U-Net were converted into segmentations containing only a single major region per image. To achieve this, the prediction maps were subjected to binary thresholding with the threshold set to the mean image intensity (Figure 4.I), and then the largest connected component (LCC) was extracted from each of these binary images ( Figure 4.II). This resulted in a set of binary liver region segmentations containing the liver dome.

Next to extract the liver dome, we applied edge detection to the LCCs/liver regions. We employed the HED model[22] used in the preprocessing stage to convert the regions to edge maps (Figure 4.III). Further refinement of these edge maps was needed to extract only the upper edges of the regions that formed the liver dome.

To aid in separating out the upper boundaries, the canny edge detection was applied to the edge maps in order, to generate edge maps with much thinner edges (Figure 4.IV). Next, morphological postprocessing with custom kernels was used to extract the liver domes from the edge maps (Figure 4.V). First erosion operations with custom kernels were performed to remove vertical edges and the lower boundaries. This was followed by the dilation operation and interpolation to correct for discontinuities in the extracted liver domes.

Evaluation

The final predicted liver dome delineations were then evaluated against the ground truth manual liver dome delineations. We used two metrics to evaluate the accuracy of the extracted liver dome delineations and thereby the overall pipeline.

The first metric is the rate of detection of the liver dome. It reported the number of images in which the pipeline was able to detect a liver dome as a percentage of the total number of images per that patient. The liver dome is considered undetectable if the resulting dome contour, after executing the post-processing steps, is non-existent or too sparse. This can be due to poor performance in the prediction of the liver dome region or issues in the post-processing steps.

The second metric is the divergence of the predicted dome from the ground truth dome in millimeters and is quantifies using the Root Mean Square Error (RMSE) values between the predicted and ground truth contours.

Results

Figure 5 illustrates the final liver dome delineations from the triggered images of three patients (with patient IDs of 01, 07 and 22 respectively). The figure details the extraction of the final liver dome delineation from the predicted liver region via postprocessing.

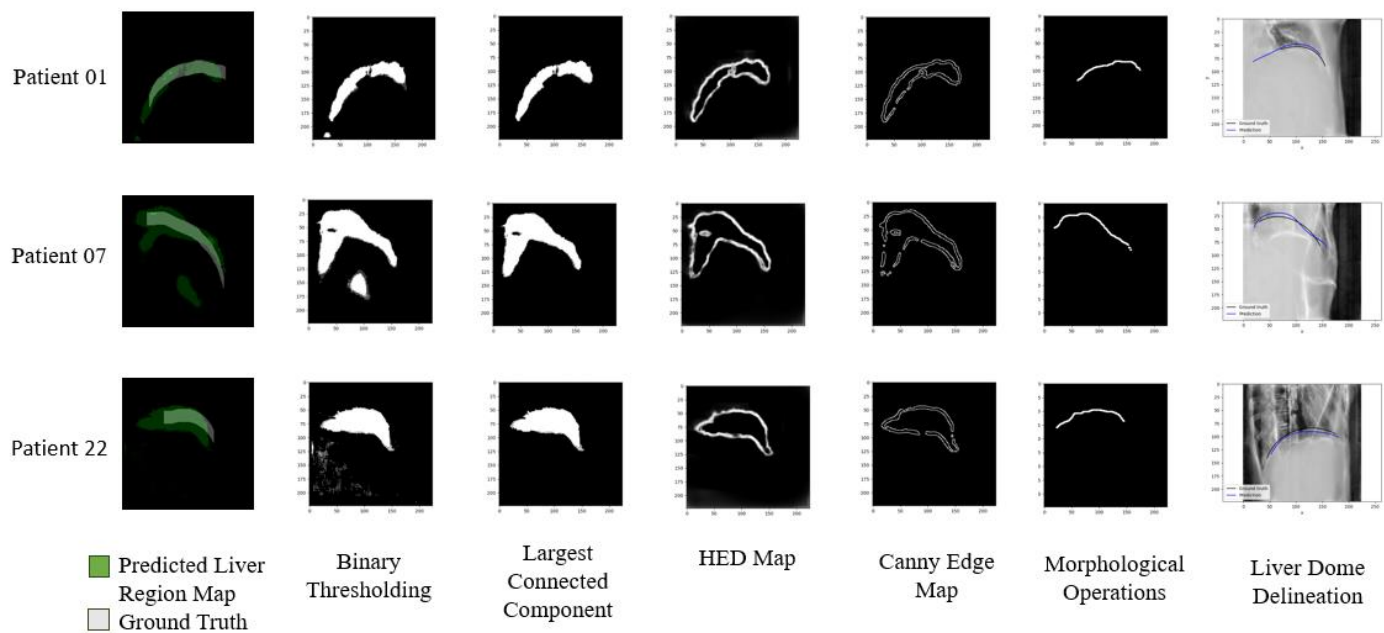

*Figure 5. Postprocessing steps to extract the liver domes from the predicted liver regions for patients 01, 07 and, 22.*

The following sections first present the quantitative evaluation of the deep learning model for liver region segmentation and then the quantitative evaluation of the overall pipeline for liver dome delineation.

Table 2 records the average IoU score of the liver dome region predictions of each patient. The IoU score which ranges from 0.0 to 1.0 is a standard metric for the evaluation of region segmentations. An IoU score of 1.0 indicates that the predicted and ground truth segmentations have complete overlap or agreement. Patients 01 to 12 (Fold$_1$) were evaluated on the deep learning model trained on patients 13 through 24 (Fold$_2$), and vice versa. The trained deep learning models achieved an IoU score greater than 0.9 for all patients.

*Table 1. Average IoU scores of predicted liver region maps for each patient.*

| Patient | Average IoU Score | Patient | Average IoU Score |
|---------|-------------------|---------|-------------------|
| 01 | 0.95 | 13 | 0.96 |
| 02 | 0.95 | 14 | 0.94 |
| 03 | 0.95 | 15 | 0.95 |
| 04 | 0.95 | 16 | 0.95 |
| 05 | 0.94 | 17 | 0.94 |
| 06 | 0.95 | 18 | 0.95 |
| 07 | 0.93 | 19 | 0.93 |
| 08 | 0.94 | 20 | 0.95 |
| 09 | 0.95 | 21 | 0.93 |
| 10 | 0.96 | 22 | 0.96 |
| 11 | 0.95 | 23 | 0.93 |
| 12 | 0.95 | 24 | 0.95 |

Figure 6 shows the training and validation learning curves for each of the two deep learning models trained on the two data folds, calculated on the metric by which the parameters of the model are being optimized- loss. In both cases, 'good' learning and generalizing of the models can be observed because the validation learning curves remain higher than the training curves, and both validation and training loss curves approach 0.1. Similarly, figure shows the IoU curves which are the performance curves for the two deep learning models. The validation IoU curves plateau at values >0.9, which shows good segmentation performance on the internal validation data split.

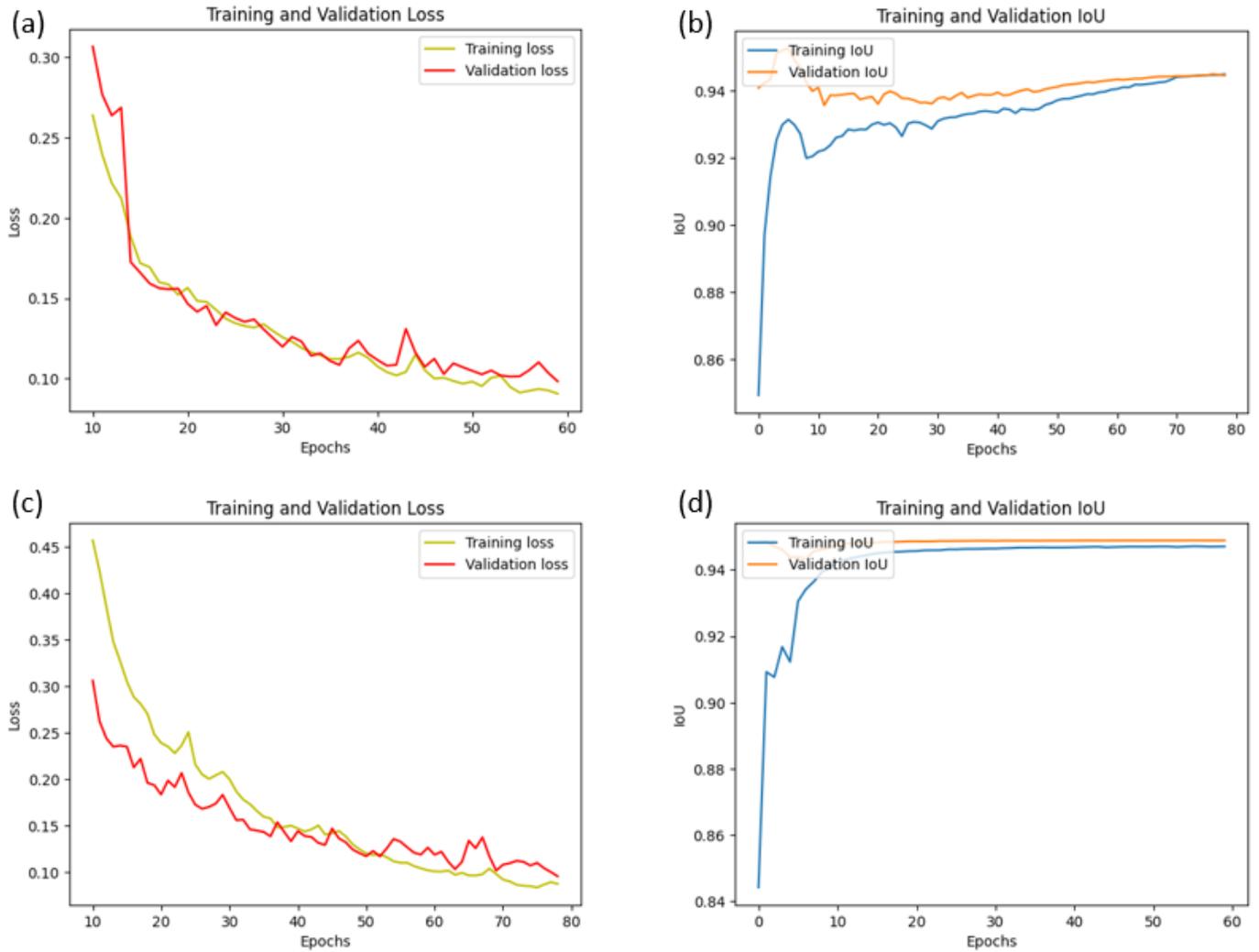

*Figure 6. (a) Training and Validation Loss curve for Fold$_1$. (b) Training and Validation IoU curve for Fold$_1$. (c) Training and Validation Loss curve for Fold$_2$. (d) Training and Validation IoU curve for Fold$_2$.*

Figure 7 (a) details the average RMSE between the automatic liver dome delineations and the ground truths for each patient. Figure 7 (b) illustrates the rate of detection of the liver domes for each patient. For patients 1 to 12, comprising Fold$_1$, the mean divergence ranged from 4.1 mm to 9.6 mm. For Fold$_2$, patients 13 to 24, this ranged from 4.2 mm to 12.0 mm. In Fold$_1$, the rate of detection fell below 85% for only patient 09 (75%). In Fold$_2$ the rate of detection for patients 13 and 15 were 61% and 59% respectively, while it ranged from 66%-91% for the rest of Fold$_2$.

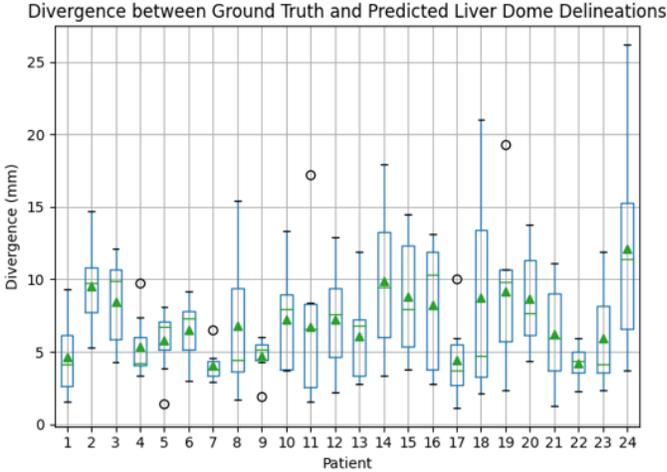
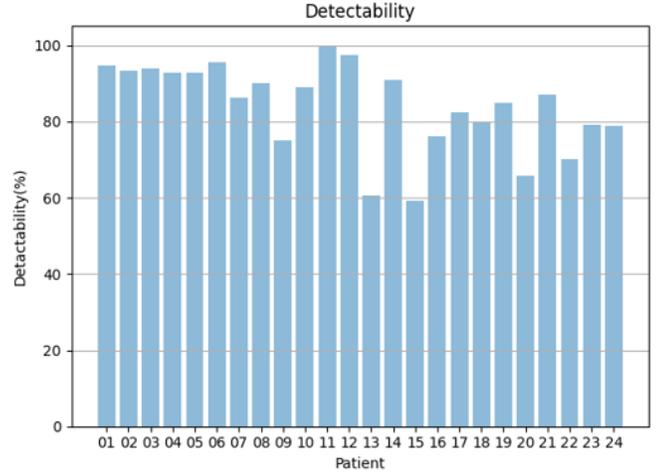

*Figure 7. (a) Scatter plot illustrating the average RMSE between the automatic liver dome delineations and the ground truths for each patient. The circles indicate outliers. (b) Bar graph illustrating the average rate of detection of the liver dome for each patient.*

Discussion

This proof-of-principle study aimed to develop a deep learning-based pipeline for the delineation of liver domes in KV triggered images. The development of such an automatic liver dome delineation technique is the initial step towards achieving automatic beam gating to improve the clinical applicability and accuracy of the online breath-hold verification technique introduced in a previous study by our institution[1].

The proposed pipeline comprises a trained deep learning model to segment the region of the liver containing the liver dome followed by morphological post-processing to extract the liver dome. The training of a deep learning model for the segmentation task on a dataset containing 500 images took approximately 30 minutes, and once trained the region segmentation and liver dome extraction takes under 1 second per triggered image. Hence, the proposed pipeline works as a fast, non-resource intensive technique for clinically feasible real-time liver dome delineation.

A key contribution of the study is the use of the two DL models-U-Net and HED. We employed the U-Net for segmentation of the liver dome region based on its success in literature for medical image segmentation. Notable community segmentation challenges, including FLARE[30], CHAOS[31], and KiTS[32], have seen successful solutions using the U-Net architecture with various modifications[33]. The U-Net, employs an encoder-decoder with skip connections[34]. Popular choices for the encoder arm include VGG16[35,36] or ResNet[37,38] pretrained on ImageNet[20], leveraging transfer learning capabilities for improved performance[21]. The idea is that the model will start out with some knowledge of natural images and will be trained or "tuned" on domain specific images for improved performance. We used this concept to our advantage with good results.

We used the HED model in two instances: for edge detection in the preprocessing stage for feature engineering and in the postprocessing stage for extracting edges from the U-Net predicted liver region. HED, based on a pretrained and modified VGGNet, outperforms conventional edge detection algorithms, successfully detecting object contours despite complexities due to image background or texture[22]. Studies, such as Rampun et al.[39], use a modified HED for breast pectoral muscle boundary segmentation, achieving Jaccard and Dice similarity scores (DSC) above 90%. Gamal et al.[40] combine Canny edge detector and HED for breast tissue classification with a resulting F1 score of 97%. Talakola et al.[41] employ an ensemble model, including the Edge U-Net model with HED, achieving a weighted DSC of 86% for intestine and stomach segmentation. Roth et al.[42] propose a two-phase modified HED approach for automatic pancreas localization and segmentation, reporting a DSC of 81%.

The performance of the deep learning model for the liver region segmentation was subjected to 2-fold cross validation and evaluated using the IoU metric. For all patients in both folds, the IoU score remained above 0.9, which indicated good segmentation performance. The training loss curves for both folds showed decreasing behavior even at the final epoch, which suggests that further learning for the models may be possible. However, the validation data while training was based on an internal 10% split, and further generalization of the models can be achieved if a hold-out validation data set can be used. We suggest this as a future work. In studies[39–42], the choice of evaluation metric for segmentation tasks is DSC (Dice Similarity Coefficient). However, we chose IoU score because it penalizes under and over segmentation tasks more than DSC and the task of the model was to predict a region that is only 20 pixels wide. Our deep learning model with its IoU scores above 0.9 is on par with the segmentation models in studies[39–42] that report evaluation metrics scoring above 80%.

The overall performance of the liver dome delineation pipeline was evaluated for each fold and all patients. Two custom metrics were introduced to evaluate liver dome delineations: the divergence in mm of the extracted liver dome from the ground truth liver dome (RMSE) averaged over all the triggered images per patient, and the rate of detection of liver domes per patient. For $Fold_1$ the average divergence and rate of detection was (6.4±1.6) mm and 91.7% respectively, and for $Fold_2$ the average divergence and rate of detection was (7.7±2.3) mm and 76.3% respectively.

Figure 8 details the liver region segmentation and subsequent liver dome extraction steps for patients 14,19, and 03, which have the lowest performance evaluation scores. It can be observed that the low divergence scores result mainly due to errors in the liver region segmentation (patient 14), failure of the morphological operations to extract the liver dome (patient 19), or both (patient 03).

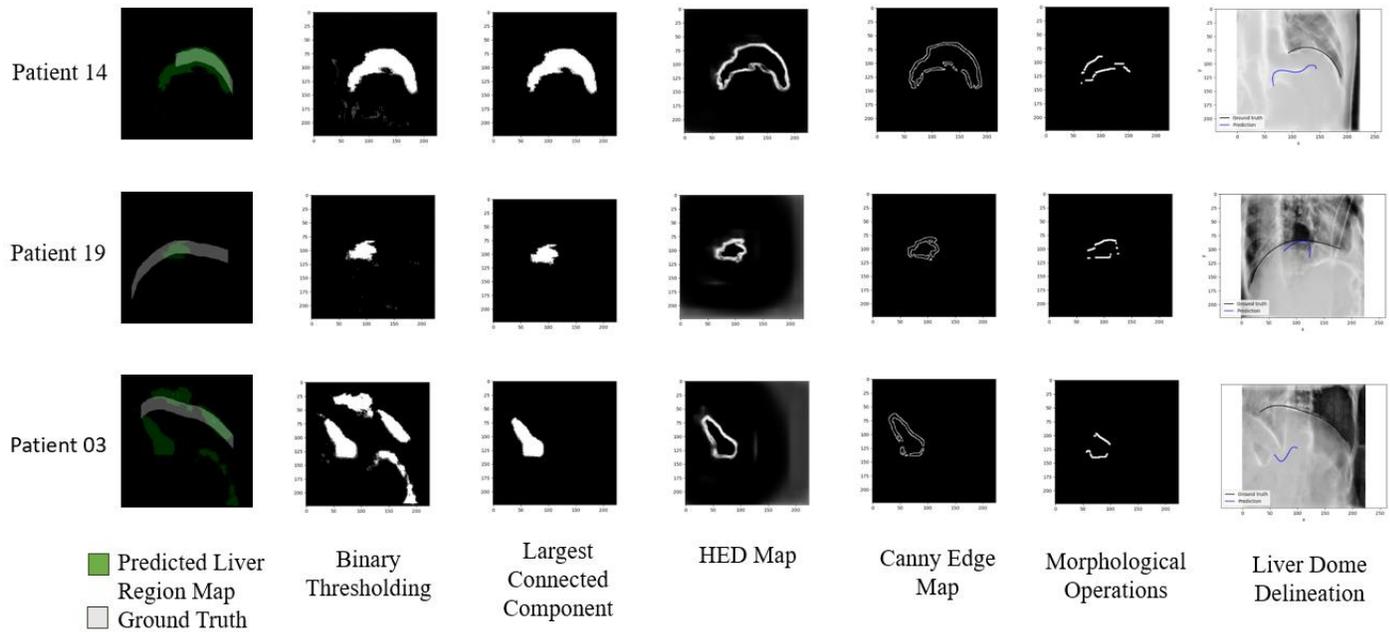

*Figure 8. Postprocessing steps to extract the liver domes from the predicted liver regions for patients 14, 19 and, 03.*

Therefore, significant improvement of the performance of the proposed pipeline may be achieved by improving the deep learning-based segmentation task, and by refining the morphological operations for increased generalizability. This will be a future direction of the study.

An additional immediate future direction of the study would be to investigate more closely the triggered images where the liver dome detection fails, to identify potential factors, such as whether the patient has had surgery etc.

A notable observation was that increasing the data size beyond 500 (to 1000 or 2000) through augmentation didn't enhance model performance and led to poorer results. It's likely that the transformations used for augmentation are insufficient to realistically model the variations in different triggered liver images, potentially introducing data bias. This would explain why increasing the data size beyond 500 did not yield better performance. A potential future direction of the study would be to better characterize and model the variations between the different triggered images.

Another limitation of this study is the comparatively small dataset used for training and validation of the proposed pipeline. Although 2-fold cross validation is used to evaluate the generalizability of the proposed method, evaluating the proposed method on an inter-institutional or multi-institutional dataset would provide a better understanding of the generalizability of the proposed pipeline.

## Conclusion

The proposed pipeline works as a proof-of-principle study for a clinically feasible, deep learning-based pipeline for automatic delineation of liver domes in kV triggered images. The incorporation of the proposed pipeline in the online breath-hold verification technique[1] could result in increased treatment accuracy and efficiency for breath-hold liver SBRT. However, further study is necessary to improve the RMSE and rate of detection to make this technique clinically relevant.